\documentclass[a4paper,fleqn]{cas-sc}

\usepackage[authoryear,longnamesfirst]{natbib}

\def\tsc#1{\csdef{#1}{\textsc{\lowercase{#1}}\xspace}}
\tsc{WGM}
\tsc{QE}
\tsc{EP}
\tsc{PMS}
\tsc{BEC}
\tsc{DE}
\begin{document}
	\let\WriteBookmarks\relax
	\def\floatpagepagefraction{1}
	\def\textpagefraction{.001}
	\shorttitle{TDMA-based scheduling with 3-egress gateway linear topology}
	\shortauthors{Linh Vu Nguyen et~al.}
	%\begin{frontmatter}
	
	\title [mode = title]{TDMA-based scheduling for multi-hop wireless sensor networks with 3-egress gateway linear topology}

	\author{Linh Vu Nguyen}
	\ead{linh.nguyen-vu760@mail.kyutech.jp}
	
	\author{Nguyen Viet Ha}
	\ead{nguyen.viet-ha503@mail.kyutech.jp}
	
	\author{Masahiro Shibata}
	\ead{shibata@cse.kyutech.ac.jp}
	
	\author{Masato Tsuru}
	\ead{tsuru@cse.kyutech.ac.jp}
	\address{Computer Science and System Engineering, Kyushu Institute of Technology, Fukuoka, Japan}
	
	\begin{abstract}
		Packet transmission scheduling on multi-hop wireless sensor networks with 3-egress gateway linear topology is studied.
		Each node generates a data packet in every one cycle period and forwards it bounded for either of gateways at edges.
		We focus on centrally-managed Time Division Multiple Access (TDMA)-based slot allocations and provide the design of a packet transmission scheduling framework with static time-slot allocation and basic redundant transmission  to reduce and recover packet losses. 
		We proposed three general path models, which would represent all possible variations of this topology, and derived a global static time-slot allocation on each of models to maximize the theoretical probability that all packets are successfully delivered to one of the gateways within one cycle period.
	\end{abstract}

	\begin{graphicalabstract}
		\includegraphics[scale=.7]{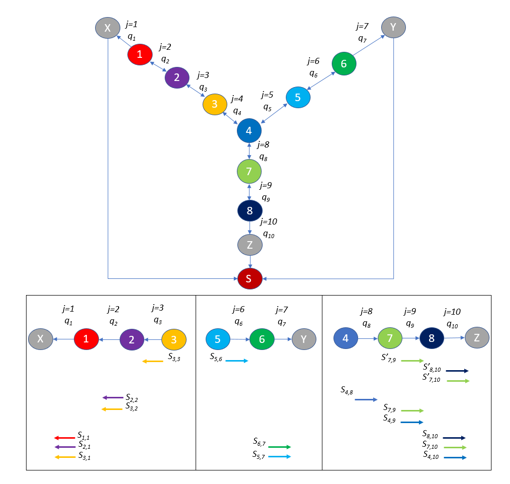}
	\end{graphicalabstract}
	
	\begin{highlights}
		\item Modeling and investigating linear topologies with three gateways at the edges (Y-shaped topology) for multi-hop wireless sensor networks to resolve interferences among relay nodes and unreliable lossy links.
		
		\item Designing a global static time-slot allocation to maximize the theoretical probability of successful delivery within one cycle period with redundant transmissions.

	\end{highlights}
	

	\maketitle

	\section{Introduction}
	
	% background and general issue
	Multi-hop wireless networks are in widespread use nowadays due to their cost-efficiency and flexibility in deployment and operation.
	They can connect nodes in an extensive coverage area larger than a single hop radio range with proper transmission power. Consequently, multi-hop wireless networks are an excellent candidate for emerging IoT systems, in which surveillance sensors are deployed along with a road, river, or electricity pylons network when a commercial communications infrastructure is unavailable or too costly.
	However, especially when the number of hops is large, multi-hop wireless networks for field monitoring often suffer from frequent packet losses due to attenuation and fading on each link as well as radio interferences of simultaneous transmissions among nodes.
	Furthermore, in typical multi-hop sensor network scenarios, since each packet conveying sensing data should be forwarded toward one of the sink nodes, the links near a sink are likely congested to forward all packets coming from upstream nodes.
	In general, to cope with frequent packet losses, 
	there is a proactive approach, e.g., redundant transmissions of original or coded packets with forwarding erasure correction (FEC); a reactive approach, e.g., retransmission of lost packets by automatic repeat request (ARQ); and a combination of them, e.g., Hybrid ARQ.
	To avoid or reduce interferences (conflicts) of simultaneous packet transmissions, 
	there is a centralized scheduling-based approach, e.g., Time Division Multiple Access (TDMA), and a decentralized contention-based approach, e.g., CSMA/CA.

	% our target
	In this paper, our target is a stationary but lossy backbone network to forward the sensing data from sensors to multiple egress ``gateways'' that are connected to a central data collection via an infrastructural reliable network.
	Therefore, a ``node'' does not represent each sensor but rather a low-cost relay node.
	Each node periodically gathers sensing data from nearby end sensors and forwards them in a hop-by-hop store-and-forward manner to one of the gateways within one cycle time-period.
	Neighboring relay nodes communicate with each other by a simple omnidirectional antenna using the same single frequency.
	Note that there must be some way to gather sensing data from end sensors connected to the relay node,  e.g., a short-range wireless link different from the links between relay nodes in terms of types and frequencies, which is out of the paper scope.

	% linear topology
	In our study scope, to be more exact, we consider kinds of stationary ``linear topologies'' 
	that are simple but typical for geographically elongated field monitoring,
	rather than mesh, complex, or dynamic topologies.
	In addition, since targeting a dedicated centrally-managed network (consisting of low-cost relay nodes, gateways, and a central server),
	we adopt a centrally-managed TDMA-based packet transmission scheduling with redundant transmissions assuming that the link layer does not provide any ARQ and transmission power adaptation mechanisms.
	A survey paper discussed the types of linear wireless sensor networks and their applications (\cite{Jawhar11}).
	In linear topologies, the number of neighboring nodes is limited as well as the distance between of neighboring nodes is generally long.
	Therefore, the number of potential interference patterns and the number of possible routing options are also limited, which may allow us to pursue an optimality easier compared with general dense topologies.
	On the other hand, a small number of possible routes will be a disadvantage in terms of robustness against a failure of nodes or links. 
	It is worth noting that we focus on cases in which gateways are placed outer-side rather than inner-side of a linear topology, e.g., at a center, which is essentially different from ``tree topologies''.
	Tree topologies are often used and can reduce the number of necessary gateways but may suffer from heavy congestions around gateways.
	In our previous work, we focused on tandemly-arranged topology networks with two gateways at both edges of a linear network (\cite{Agus16, Kimura20, Yoshida20}).
	Assuming that the topology, the data transmission rate (i.e., bandwidth) and the time-averaged packet loss rate of each link, the packet size, and the packet generation rate of each node are known, 
	we successfully derived an ``optimal'' static packet transmission time-slot allocation under a basic controlled redundant transmission scheme.

	\begin{figure}[h]
		\centering
		{
			\includegraphics[scale=.45]{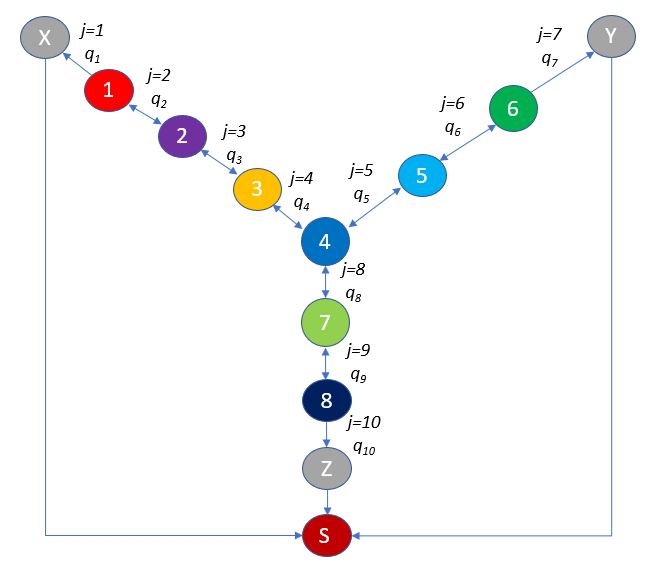}
			\caption{An example of Y-shaped network topology used in this research}
			\label{jf02-Y-shap}
		}
	\end{figure}

	Following our previous work, in this paper, we present the models and the performance investigation on the 3-egress gateway linear topology that is a linear topology with three gateways at the edges, called the Y-shaped topology. In any Y-shaped topology, there is a central node that potentially has three links but some of three links are not necessarily used for data transmission.
	An example is illustrated in Fig.~\ref{jf02-Y-shap}.
	The Y-shaped topology represents not only a topological form as a graph but three gateways' locations at the edges (outer-side) as mentioned above.
	This paper is a fully-extended version of our most recent publication in which we used the term T-shaped topology (\cite{Nguyen20}).
	Our proposed scheme aims at a global static time-slot allocation on Y-shaped topologies to maximize the theoretical probability that all packets are successfully delivered to one of the gateways within one cycle period with redundant transmissions.
	A difficulty of transmission scheduling on Y-shaped topologies is the existence of patterns of potential interferences around the central node, in contrast to tandemly-arranged linear topologies with two gateways at the edges.
	On the other hand, such interferences may be efficiently avoidable because the data forwarding directions are opposite, in contrast to tree-like topologies with a gateway at the center.
	Note that, in our scheme, a central management server is assumed to be able to know or estimate necessary information such as the packet loss rate of each link, compute a global time-slot allocation, and deliver the derived schedule to each node. 
	Such system implementation issues will be discussed later in the discussion section because it is not the main scope of this paper.

	The rest of this paper is organized as follows.
	The related work is reviewed in Sec.~2.
	The Y-shaped topology and the path models are defined in Sec.~3.
	Section 4 explains how to derive optimal time-slot allocations by using an example topology, which is evaluated through numerical simulations in Sec.~5. 
	Discussions are provided in Sec.~6 and finally, Sec.~7 concludes the paper.
	The appendix is given to explain an example of the detailed derivation of the mathematical formula.

	\section{Related work}
	
	For multi-hop wireless networks, there have been a variety of studies devoted to cope with the lossy unreliable wireless radio links and the conflicts (interferences) among simultaneous transmissions on adjacent links depending on a variety of requirements and restrictions.
	Even in TDMA-based transmission scheduling to resolve two fundamental bottlenecks of multi-hop wireless networks, various methods have been developed (\cite{Sgora15}). In theory, it is conducted by defining a conflict-free TDMA for a given set of links, which is formulated as a graph coloring. In addition, wireless conflicts are able to model with conflict graphs (\cite{Jain05}, \cite{Ramanathan99}). As a result, for example, \cite{Jain05} obtained a graph coloring on the conflict graph, a conflict-free schedule formed from independent sets with appropriate cardinality. 
	In specific adjacent links, some studies addressed these issues by utilizing distributed implementation of RAND, a randomized time slot scheduling algorithm (\cite{Rhee09}). Besides, \cite{Ergen10} introduced the shortest schedules based on two centralized algorithms, which was considered a more facile method to evaluate the performance of distributed algorithms. In a similar concern about the shortest schedule,  \cite{sasaki16} introduced a min-max model and a min-sum model as an efficient method for this point, but their large-scale application was hindered. \cite{Zeng14} utilized scheduling algorithm, which was developed from the collaboration of nodes, to improve the packet receive ratio and energy efficiency.

	However, almost all these studies concentrated on developing conflict graphs or heuristics to avoid interference of simultaneous data transmission on general network topology, and they do not deal with optimal redundant transmissions to recover lost packets.On the other hand, our previous work (\cite{Agus16, Kimura20, Yoshida20}) in this theme provided a packet transmission scheduling framework restricted to  tandemly-connected topologies onlys. 
	\cite{Nguyen20} provided a preliminary result on the T-shaped topology with 3 gateways but not offered a comprehensive study. Herein, nodes located near the central point have the influence on interference following 3 directions of topology. Therefore, they designed a schedule of when and to which gateway each node sends a packet. It is vital to control the efficiency of the system.

	\section{The Y-shaped topology model}
	
	\begin{figure}[h]
		\centering
		\includegraphics[scale=.65]{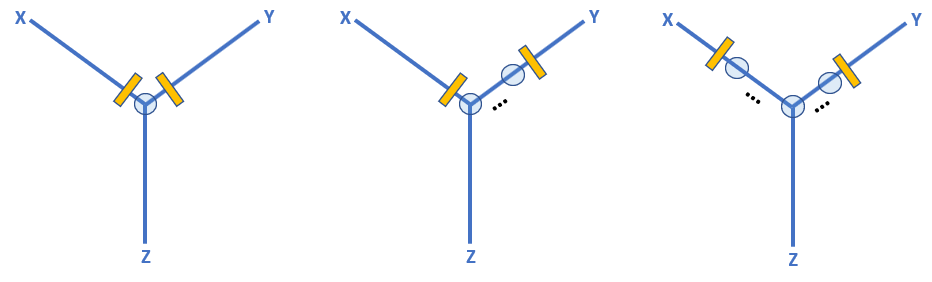}
		\caption{Y-shaped network topology model}
		\label{jf01-three model}
	\end{figure}	
	
	In a Y-shaped topology network,
	the nodes and links are numbered separately (starting from 1) as shown in Fig.~\ref{jf02-Y-shap}.
	The packet loss rate of link $j$ is denoted as $q_{j} \:(0 < q_{j} < 1)$, and the packet generation rate of node $i$ is denoted as positive integer $r_{i}$. 
	As explained in Sec.~1, each node considered here is a relay node of the Y-shaped backbone of a sensor network.
	Hence the packet generation rate of a node represents the number of packets to convey the total amount of sensing data in one cycle period of $D$ gathered from end sensors managed by the node.
	In other words, node $i$ is assumed to generate $r_{i}$ packets at (or before) the beginning of each $D$ and those packets are forwarded toward a gateway either $X$, $Y$, or $Z$.
	For concise formulations, all packets are assumed to have the same size and all links are to have the same data transmission rate. Therefore, let $U$ be the time duration of one time-slot, i.e., one packet can be transmitted on a link between adjacent two nodes in $U$ unit time, Then the total number $T$ of slots in one cycle period is equal to $D/U$.

	We assume that each packet is forwarded along a single path (not a multi-path) toward only one of the three gateways.
	Based on this assumption, this study models the Y-shaped topology in three types as shown in Fig.~\ref{jf01-three model}.
	A separation link is a link on which no packet is transmitted because the two nodes at both sides of the link send packets in opposite directions, i.e., the link is between two nodes each of which is at the most upstream of a path.
	Since the Y-shaped topology has a single central node and three branches (called ``segment'') terminated by gateways X, Y, and Z,
	there should be two separation links that separate the topology into three paths with different numbers of nodes in general.
	We call it ``the path model'' representing packet routing paths; different path models can be considered by choosing the locations of the separation links. 
	To be more exact, we can assume one separation link is in the segment to X and the other is in the segment to Y; no separation link in segment to Z without loss of generality.
	In this sense, the gateway name $X$, $Y$, and $Z$ should be given depending on the path model.
	The locations of separation links can classify all nodes into three groups called $S_X$, $S_Y$, and $S_Z$ which are the set of nodes whose packets are forwarded to X, Y, and Z, respectively.
	It is called the $l$-$r$-$d$ model where $l$, $r$, and $d$ represent the number of nodes in $S_X$, $S_Y$, and $S_Z$, respectively.
	Furthermore, based on the locations of separation links, three types of path models are considered. 
	Type 1: no node on segment to X and Y is in $S_Z$.
	Type 2: some nodes on only one of segments to X and Y are in $S_Z$.
	Type 3: some nodes on segment to X and also some nodes in segment to Y are in $S_Z$.
	
\begin{figure}
		\centering
		\includegraphics[scale=.55]{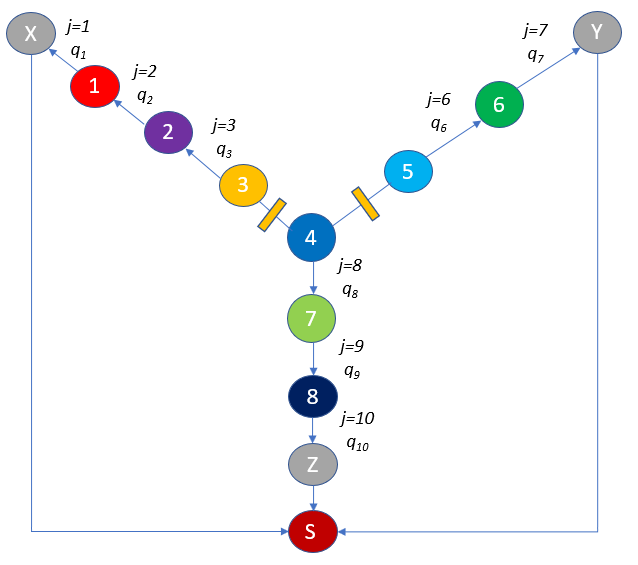}
		\caption{The 3-2-3 path model in the network topology}
		\label{jf03-m323}
	\end{figure}

	Those types would represent all possible variations of Y-shaped topologies and affect the potential interference patterns among nodes nearby the central node that strongly impact the design of slot allocations, i.e., slot allocation patterns.
	For example, as shown in Fig.~\ref{jf03-m323}, if the first separation link is set between nodes 3 and 4, and the second is set between nodes 4 and 5, then the path model is 3-2-3 model and this is Type 1. Group $S_X$ has 3 nodes whose packets are forwarded to gateway X ($S_X = \{1,2,3\}$), group $S_Y = \{5,6\}$, and group $S_Z = \{4,7,8\}$. 
	Similarly, Fig.~\ref{jf06} and Fig.~\ref{jf09} illustrate the 2-2-4 model of Type 2 and 2-1-5 model of Type 3, respectively.

	%%%%%%%%%%%%%%%%%%%
	
	\section{Path models and Time-slot allocation}
	
	In designing a global static time-slot allocation, there are two issues: how much each node can utilize a limited number $T$ of time-slots with redundant packet transmissions by considering the upstream-downstream relationship among nodes and the packet loss rate of each link; and how much it can avoid radio interferences in simultaneous transmissions.
	To solve those issues, the following steps are performed.
	First, we list all possible path models on a given Y-shaped topology.
	Second, for each path model, by considering the potential interference patterns around the central node, we list all possible ``slot allocation patterns''.
	To prohibit nearby nodes from harmful simultaneous transmissions, a static interference avoidance policy based on the distance between nodes is adopted.
	Although the possible patterns are limited, we need to use geographical information such as distances and environmental conditions which are not represented by an abstract topology.
	For each pattern of each path model, we derive a static time-slot allocation to maximize the theoretical probability that all packets are successfully delivered to gateways within the total time-slots of $T$.
	Lastly, by comparing all results in terms of the maximum theoretical probability of successful delivery, we can select the best path model with the best slot allocation.

	We explain an optimal static time-slot allocation for each allocation pattern of each path model on an example topology with $8$ nodes illustrated in Fig.~\ref{jf02-Y-shap}.
	To express a slot allocation, in general, we denote $s_{i,j,k}$ as the number of slots allocated, i.e., available to use, for the $k$-th packet generated by node $i$ on link $j$ in each cycle period.
	In other words, each node redundantly transmits a possessed packet (that is originally the $k$-th packet generated by node $i$) on downstream link $j$ in $s_{i,j,k}$ times. 
	If a packet is lost somewhere in upstream between that node and the node which generated the packet, the slots allocated to the lost packet are used for the next packet.
	However, for concise explanation, packet generation rate $r_i$ is assumed to be $1$ and thus $s_{i,j}$ is used instead of $s_{i,j,k}$.
	The extension to heterogeneous packet generation rates $\{r_i\}$ is somewhat straight-forward as shown in Appendix.
	Note that we also introduce $s'_{i,j}$ to indicate the number of slots for an early stage transmission which happens before or at the same time of transmission of the most upstream node in the path, i.e.,
	a concurrent use of the same time-slots for the same direction transmissions to increase the efficiency.

	In general, the maximization problem for optimal slot allocations is defined as follows.
	The success probability of delivery of a packet generated by node $i$ is denoted by $M_i(\mathsf{s_i})$ that can be calculated by a slot allocation $\mathsf{s_i} = \{s_{i,j}, s'_{i,j}|j=\ldots\}$ for $i=1,2,\ldots,8$.
	Hence the problem to solve is
	\begin{eqnarray*}
		\max \:M(\mathsf{s}) &=& \prod_{i=1}^{8}M_{i}(\mathsf{s_i}) 
		\:\:\: \mbox{subject to}
		\nonumber
		\\
		T &=& (\mbox{a linear function of}\: \mathsf{s}\: \mbox{in group X})
		\nonumber
		\\
		T &=& (\mbox{a linear function of}\: \mathsf{s}\: \mbox{in group Y})
		\nonumber
		\\
		T &=& (\mbox{a linear function of}\: \mathsf{s}\: \mbox{in group Z})
		\label{maxs_m323p1}
	\end{eqnarray*}
	where $\mathsf{s} = \{s_{i,j}, s'_{i,j}|i=1,2,\ldots,8; \:j=\ldots\}$.

	\subsection{Slot allocation for 3-2-3 model}
	
	Fig.~\ref{jf03} shows the 3-2-3 path model. 
	In this example, since we assume nodes 3 and 7 are in the radio propagation distance, 3 and 4 cannot send at the same time (to avoid interference at node 7). Since nodes 5 and 7 are in the propagation distance, 5 and 4 cannot send at the same time  (to avoid interference at node 7). 
	On the other hand 3, 5, and 7 can send to their next node at the same time.
	We have two patterns for slot allocations. In pattern 1, we prioritize the transmission in groups $S_X$ (node 3-2-1) and $S_Y$ (node 5-6) first, then group $S_Z$ (node 4-7-8). In pattern 2, we prioritize group $S_Z$ first, then groups $S_X$ and $S_Y$.
	In other words, in pattern 2, the most upstream side node toward Z (i.e., node 4) can start its transmission earlier than the most upstream nodes toward X and Y (i.e., nodes 3 and 5).

	\begin{figure}[b]
		\centering	
		\includegraphics[scale=.55]{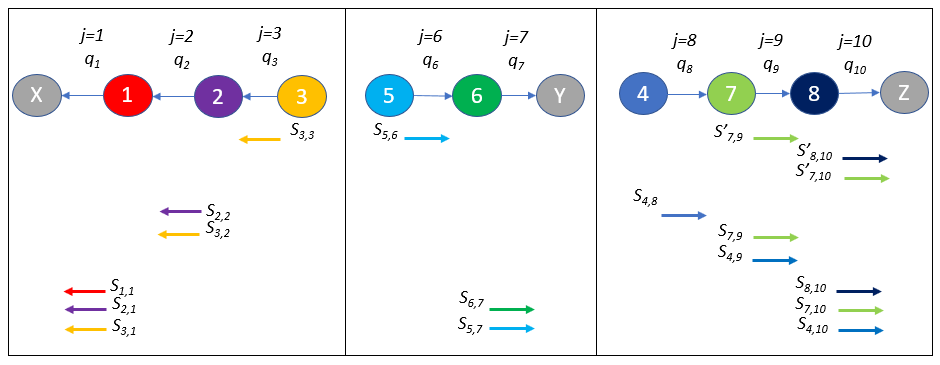}
		\caption{Transmission scheduling on 3-2-3 model (pattern 1, all $r_i$=1)}
		\label{jf04-m323-p1}
	\end{figure}
	
	%%%%%%%%%%%%%%%%Model 3-2-3 Pattern 1

	Pattern 1 is adopted 
	when node group $S_X$ or $S_Y$ is a bottleneck.
	We solve a sub-problem for group $S_X$ and $S_Y$ first.
	Fig.~\ref{jf04-m323-p1} shows slot allocations pattern 1 in 3-2-3 model.
	In group $S_X$,
	\begin{eqnarray*}
		M_1&=&(1-q_{1}^{{s_{1,1}}}),
		\nonumber
		\\
		M_2&=&(1-q_{1}^{{s_{2,1}}})(1-q_{2}^{s_{2,2}})
		\nonumber
		\\ 
		M_3&=&(1-q_{1}^{s_{3,1}})(1-q_{2}^{s_{3,2}})(1-q_{3}^{s_{3,3}})
		\label{M_3_t323p1}
	\end{eqnarray*}
	In group $S_Y$,
	\begin{eqnarray*}
		M_5&=&(1-q^{s_{5,6}}_6)(1-q^{s_{5,7}}_7)
		\nonumber
		\\
		M_6&=&(1-q_{7}^{{s_{6,7}}})
		\label{M6_t323p1}
	\end{eqnarray*}
	In group $S_Z$, $s_{7,9}$ and $s'_{7,10}$ cannot be 0 at the same time, in any optimal schedule.
	\begin{eqnarray*}
		M_8&=&(1-q_{10}^{{s_{8,10}}+{s'_{8,10}}}),
		\nonumber
		\\
		M_7&=&
		\left\{
		\begin{array}{lll}
			(1-q_{10}^{{s'_{7,10}}+{s_{7,10}}})(1-q_{9}^{s'_{7,9}}) 
			& \mbox{if } & s_{7,9} = 0,
			\\
			(1-q_{10}^{{s_{7,10}}})(1-q_{9}^{{s'_{7,9}}+{s_{7,9}}}) 
			& \mbox{if}  & s'_{7,10} = 0,
		\end{array}
		\right.
		\nonumber
		\\ 
		M_4&=&(1-q_{10}^{s_{4,10}})(1-q_{9}^{s_{4,9}})(1-q_{8}^{s_{4,8}})
		\label{M_8_t323p1}
	\end{eqnarray*}

	To get a slot allocation in group $S_X$
	that maximizes $M_{1}M_{2}M_{3}$ subject to 
	\begin{eqnarray}
	T=s_{1,1}+s_{2,1}+s_{2,2}+s_{3,1}+s_{3,2}+s_{3,3},
	\label{max123_t1p1}
	\end{eqnarray}
	we apply the Lagrangian multiplier to a relaxation version of this problem to derive equations (\ref{s11_t323p1}) where ${s_{i,j}}$ are not restricted to natural numbers and $\alpha$ is an unknown adjunct variable; please see Appendix.
	\begin{eqnarray}
	s_{1,1}=s_{2,1}=s_{3,1}=-\frac{\log(1-\alpha\log(q_{1}))}{\log(q_{1})}
	\nonumber
	\\
	s_{2,2}=s_{3,2}=-\frac{\log(1-\alpha\log(q_2))}{\log(q_2)}, \quad s_{3,3}=-\frac{\log(1-\alpha\log(q_3))}{\log(q_3)}
	\label{s11_t323p1}
	\end{eqnarray}
	From Eqs.(\ref{max123_t1p1}) and (\ref{s11_t323p1}),
	$\alpha$ can be numerically solved to get the real number solution $\{s_{i,j}\}$ of the relaxed problem.
	Then we should seek an appropriate natural number solution as the number of allocated slots near the derived real number solution.
	Let $\{s_{i,j}^*\}$ be the natural number solution obtained for $S_X$;
	let $\mathbf{a_3}$ be $s_{3,3}^*$.

	Independently and similarly to $S_X$,
	to get a slot allocation in group $S_Y$
	that maximizes $M_{5}M_{6}$ subject to 
	\begin{eqnarray}
	T&=&s_{5,6}+s_{5,7}+s_{6,7},
	\label{max56_t323p1}
	\end{eqnarray}
	we have 
	\begin{eqnarray}
	s_{5,7}=s_{6,7}=-\frac{\log(1-\beta\log(q_7))}{\log(q_7)}, \quad s_{5,6}=-\frac{\log(1-\beta\log(q_6))}{\log(q_6)}
	\label{s5657p1}
	\end{eqnarray}
	where $\beta$ is an unknown adjunct variable.
	From Eqs.(\ref{max56_t323p1}) and (\ref{s5657p1}),
	$\beta$ can be numerically solved to get the real number solution $\{s_{i,j}\}$ of the relaxed problem.
	Then
	we obtain the natural number solution $\{s_{i,j}^*\}$ for $S_Y$.
	Let $\mathbf{a_6}$ be $s_{5,6}^*$.

	Finally, to find a slot allocation in group $S_Z$
	using $\mathbf{a} = \max(\mathbf{a_3}, \mathbf{a_6})$, 
	we tentatively maximize $M_{4}M_{7}M_{8}$ without considering $s'_{8,10}$, $s'_{7,10}$, $s'_{7,9}$ subject to 
	\begin{eqnarray}
	T=s_{8,10}+s_{7,9}+s_{7,10}+s_{4,8}+s_{4,9}+s_{4,10}
	\label{T123_m323p1}
	\end{eqnarray}
	where its solution $\{s_{i,j}^*\}$ can be solved in the same way.
	Let $\mathbf{b}_8$ be $s_{4,8}^*$, $\mathbf{b}_9$ be $s_{4,9}^*$, $\mathbf{b}_{10}$ be $s_{4,10}^*$.
	There are five cases:
	(c1) $\mathbf{b}_9 \geq \mathbf{a}$; (c2) $\mathbf{b}_9 < \mathbf{a}$ and $\mathbf{b}_{10} \geq \mathbf{a}$;
	(c3) $\mathbf{b}_9+\mathbf{b}_{10} \geq \mathbf{a}$;
	(c4) $\mathbf{b}_9 + 2\mathbf{b}_{10} \geq \mathbf{a}$;
	(c5) $\mathbf{b}_9+2\mathbf{b}_{10} < \mathbf{a}$.
	\begin{itemize}
		\item
		In (c1), a final natural number solution is:
		\begin{eqnarray*}
			s_{4,8}=\mathbf{b}_8 , \quad s'_{7,9}=\mathbf{a} , \quad s_{7,9}=\mathbf{b}_9-\mathbf{a} , \quad s_{4,9}=\mathbf{b}_9
			\\
			s_{8,10}=s_{7,10}=s_{4,10}=\mathbf{b}_{10}, \quad s'_{8,10}=s'_{7,10}=0
			\label{X_case1_m323p1}
		\end{eqnarray*}
		
		\item
		In (c2), a final natural number solution is:
		\begin{eqnarray*}
			s_{4,8}=\mathbf{b}_8 , \quad s'_{7,9}=0 , \quad s_{7,9}=s_{4,9}=\mathbf{b}_9 ,
			\\
			s_{8,10}=\mathbf{b}_{10} - \mathbf{a},\quad s_{7,10}=s_{4,10}=\mathbf{b}_{10}, \quad s'_{8,10}=\mathbf{a},\quad  s'_{7,10}=0
			\label{X_case2_m323p1}
		\end{eqnarray*}
		
		\item
		In (c3), a final natural number solution is:
		\begin{eqnarray*}
			s_{4,8}=\mathbf{b}_8 , \quad s'_{7,9}=\mathbf{b}_9 , \quad s_{7,9}=0, s_{4,9}=\mathbf{b}_9 , \quad s'_{8,10}=\mathbf{a}-\mathbf{b}_9,
			\\
			s_{8,10}=\mathbf{b}_{10} - (\mathbf{a}-\mathbf{b}_{9}),\quad  s_{7,10}=s_{4,10}=\mathbf{b}_{10},\quad  s'_{7,10}=0
			\label{X_case3_m323p1}
		\end{eqnarray*}	
		
		\item
		In (c4), a final natural number solution is:
		\begin{eqnarray*}
			s_{4,8}=\mathbf{b}_8 , \quad s'_{7,9}=\mathbf{b}_9 , \quad s_{7,9}=0,\quad  s_{4,9}=\mathbf{b}_9 , \quad s'_{8,10}=\mathbf{b}_{10},\quad  s_{8,10}=0,
			\\
			s'_{7,10}=  \mathbf{a}-(\mathbf{b}_{10}+\mathbf{b}_{9}),\quad  s_{4,10}=\mathbf{b}_{10},\quad  s_{7,10}=\mathbf{b}_{10}-(\mathbf{a}-(\mathbf{b}_{10}+\mathbf{b}_{9}))
			\label{X_case4_m323p1}
		\end{eqnarray*}	
		
		\item
		Case (c5) requires to solve other two equations independently:
		\begin{itemize}
			\item[(i)]
			For nodes $7$ and $8$, the time-slot region is $\mathbf{a}$.
			\begin{eqnarray*}
				\mathbf{a}&=&s_{7,9}+s_{7,10}+s_{8,10}
				\label{X_case5_1_m323p1}
			\end{eqnarray*}
			By letting $s^*_{i,j}$ be its solution,  a final natural number solution is:
			\begin{eqnarray*}
				s'_{7,9}=s^*_{7,9}, \quad s_{7,9}=0 , \quad s'_{7,10}=s_{7,10}=s^*_{8,10}\quad s_{7,10}=s_{8,10}=0
				\label{X_case51_m323p1}
			\end{eqnarray*}
			
			\item[(ii)]	
			For node 4, the time-slot region is $T-\mathbf{a}$.
			\begin{eqnarray*}
				T-\mathbf{a}&=&s_{4,8}+s_{4,9}+s_{4,10}
				\label{X_case5_3_m323p1}
			\end{eqnarray*}
			By letting $s^{**}_{i,j}$ be its solution, a final natural number solution is:
			\begin{eqnarray*}
				s_{4,8}=s^{**}_{4,8}, \quad s_{4,9}=s^{**}_{4,9} , \quad s_{4,10}=s^{**}_{4,10}
				\label{X_case52_m323p1}
			\end{eqnarray*}
			
		\end{itemize}
	\end{itemize}
	
	%%%%%%%%%%%%%%%%
	
	\begin{figure}[b]
		\centering
		\includegraphics[scale=.55]{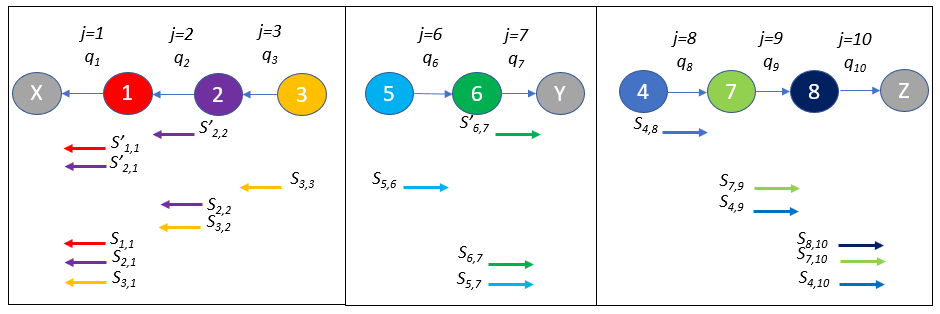}
		\caption{Transmission scheduling on 3-2-3 model (pattern 2, all $r_i$=1)}
		\label{jf05-m323-p2}
	\end{figure}

	Pattern 2 is adopted 
	when node group $S_Z$ is a bottleneck.
	We solve a sub-problem for group $S_Z$ first.
	Fig.~\ref{jf05-m323-p2} shows slot allocations pattern 2 in 3-2-3 model.
	In group $S_X$, $s_{2,2}$ and $s'_{2,1}$ cannot be 0 at the same time, in any optimal schedule.
	\begin{eqnarray*}
		M_1&=&(1-q_{1}^{{s_{1,1}}+{s'_{1,1}}}),
		\nonumber
		\\
		M_2&=&
		\left\{
		\begin{array}{lll}
			(1-q_{1}^{{s'_{2,1}}+{s_{2,1}}})(1-q_{2}^{s'_{2,2}})
			& \mbox{if } & s_{2,2} = 0,
			\\
			(1-q_{1}^{{s_{2,1}}})(1-q_{2}^{{s'_{2,2}}+{s_{2,2}}})
			& \mbox{if}  & s'_{2,1} = 0,
		\end{array}
		\right.
		\nonumber
		\\ 
		M_3&=&(1-q_{1}^{s_{3,1}})(1-q_{2}^{s_{3,2}})(1-q_{3}^{s_{3,3}})
		\label{M_3_t323p2}
	\end{eqnarray*}
	In group $S_Y$,
	\begin{eqnarray*}
		M_5&=&(1-q^{s_{5,6}}_6)(1-q^{s_{5,7}}_7)
		\nonumber
		\\
		M_6&=&(1-q^{{s_{6,7}}+{s'_{6,7}}}_7)
		\label{M6_t323p2}
	\end{eqnarray*}
	In group $S_Z$,
	\begin{eqnarray*}
		M_4&=&(1-q_{8}^{s_{4,8}})(1-q_{9}^{s_{4,9}})(1-q_{10}^{s_{4,10}})
		\nonumber
		\\
		M_7&=&(1-q_{9}^{s_{7,9}})(1-q_{10}^{s_{7,10}})
		\nonumber
		\\
		M_8&=&(1-q_{10}^{s_{8,10}})
		\label{M_8_t323p2}
	\end{eqnarray*}

	The following process is almost the same approach as Pattern 1.
	To get a slot allocation in group $S_Z$
	that maximizes $M_{4}M_{7}M_{8}$ subject to 
	\begin{eqnarray*}
		T=s_{8,10}+s_{7,9}+s_{7,10}+s_{4,8}+s_{4,9}+s_{4,10},
		\label{max478_t1}
	\end{eqnarray*}
	we have 
	\begin{eqnarray*}
		s_{4,10}=s_{7,10}=s_{8,10}=-\frac{\log(1-\gamma\log(q_{10}))}{\log(q_{10})}
		\nonumber
		\\
		s_{4,9}=s_{7,9}=-\frac{\log(1-\gamma\log(q_9))}{\log(q_9)}, \quad s_{4,8}=-\frac{\log(1-\gamma\log(q_8))}{\log(q_8)}
		\label{s49_t323}
	\end{eqnarray*}
	where $\gamma$ can be numerically solved to get the real number solution of the relaxed problem, and then obtain the natural number solution $\{s_{i,j}^*\}$ for $S_Z$.

	The obtained solution $s_{4,8}^*$ for $S_Z$ is used
	to solve group $S_Y$.
	By starting from the tentative maximization of $M_{5}M_{6}$ without considering $s'_{6,7}$ subject to 
	\begin{eqnarray*}
		T&=&s_{5,6}+s_{5,7}+s_{6,7},
		\label{max56_t323}
	\end{eqnarray*}
	a final natural number solution
	$(s_{5,6}^*, s_{5,7}^*, s_{6,7}^*, (s'_{6,7})^*)$
	is obtained after checking a few conditions (cases) in a similar manner as the pattern 1.

	The natural number solution $s_{4,8}^*$ for is also used
	to solve group $S_X$.
	By starting from the tentative maximization of
	$M_{1}M_{2}M_{3}$ without considering $s'_{1,1}$, $s'_{2,1}$, $s'_{2,2}$ subject to 
	\begin{eqnarray*}
		T=s_{1,1}+s_{2,1}+s_{2,2}+s_{3,1}+s_{3,2}+s_{3,3},
		\label{T123_m323}
	\end{eqnarray*}
	a final natural number solution
	$(s_{3,3}^*, s_{3,2}^*, s_{2,2}^*, (s'_{2,2})^*, s_{3,1}^*, s_{2,1}^*, s_{1,1}^*, (s'_{2,1})^*, (s_{1,1})^*)$
	is obtained after checking a few conditions.

	\subsection{Static slot allocation for 2-2-4 model}
	
\begin{figure}
		\centering
		\includegraphics[scale=.45]{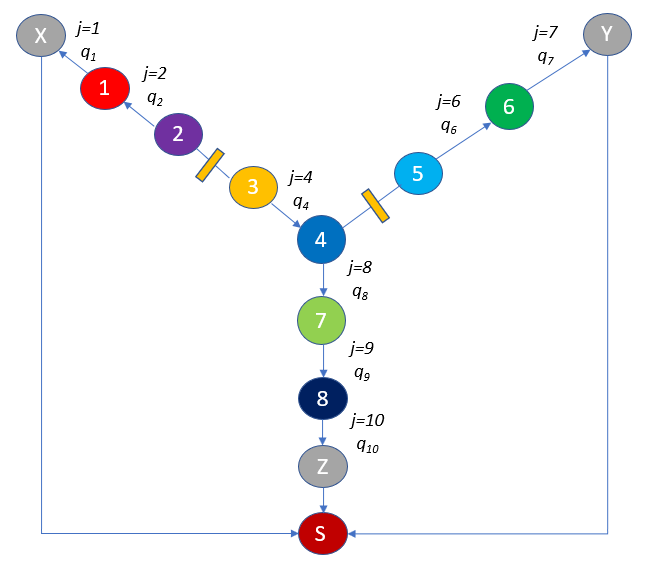}
		\caption{The 2-2-4  path model in the network topology}
		\label{jf06-m224}
	\end{figure}

	Fig.~\ref{jf06-m224} shows the 2-2-4 path model.
	In this example, since nodes 4 and 5 are assumed to be in the radio propagation distance, nodes 3 and 5 cannot send at the same time (to avoid an interference at node 4). Similarly, since nodes 5 and 7 are in the propagation distance, nodes 4 and 5 cannot send at the same time  (to avoid an interference at node 7). 
	On the other hand, 5 and 7 can send to its next node at the same time.
	We have two patterns for slot allocations. In pattern 1, we prioritize the transmission in group $S_Z$ (node 3-4-7-8) first, then group $S_Y$ (node 5-6). In pattern 2, we prioritize group $S_Y$ first, then group $S_Z$.
	Note that group $S_X$ is independent and solved separately.

	%%%%%%%%%%%%%%%%  Model 2-4-2 Pattern 1
	\begin{figure}[b]
		\centering
		\includegraphics[scale=.55]{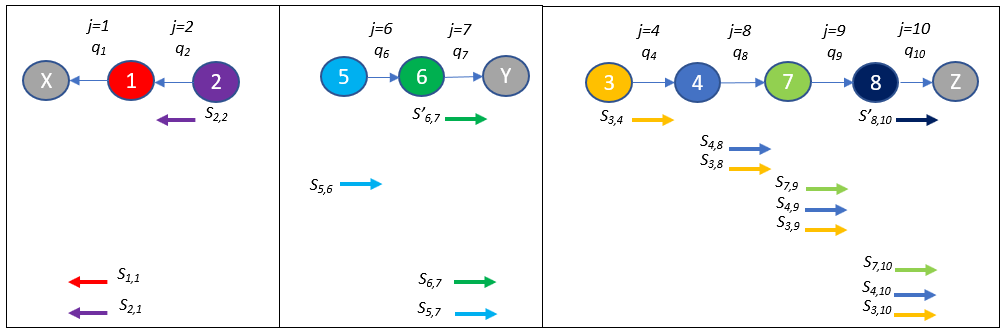}
		\caption{Transmission scheduling on 2-2-4 model (pattern 1, all $r_i$=1)}
		\label{jf07-m224-p1}
	\end{figure}

	Pattern 1 is adopted 
	when node group $S_Z$ is a bottleneck.
	We solve a sub-problem for group $S_Z$ first.
	Fig.~\ref{jf07-m224-p1} shows slot allocation pattern 1 in the 2-2-4 model. 
	In group $S_X$,
	\begin{eqnarray*}
		M_1&=&(1-q_{1}^{s_{1,1}}),\quad
		M_2\:=\:(1-q_{1}^{s_{2,1}})(1-q_{2}^{s_{2,2}})
		\label{M_12_t224p1}
	\end{eqnarray*}
	In group $S_Z$,
	\begin{eqnarray*}
		M_3&=&(1-q_{4}^{s_{3,4}})(1-q_{8}^{s_{3,8}})(1-q_{9}^{s_{3,9}})(1-q_{10}^{s_{3,10}})
		\nonumber
		\\	
		M_4&=&(1-q_{8}^{s_{4,8}})(1-q_{9}^{s_{4,9}})(1-q_{10}^{s_{4,10}})
		\nonumber	
		\\
		M_7&=&(1-q^{s_{7,9}}_9)(1-q^{s_{7,10}}_{10}), \quad
		M_8\:=\:(1-q^{s'_{8,10}}_{10})
		\label{M3478_m224p1}
	\end{eqnarray*}
	In group $S_Y$,
	\begin{eqnarray*}
		M_5&=&(1-q_{6}^{s_{5,6}})(1-q_{7}^{s_{5,7}}), \quad
		M_6\:=\:(1-q_{7}^{{s_{6,7}}+{s'_{6,7}}})
		\label{M_56_m224p1}	
	\end{eqnarray*}	
	
	The process to get a slot allocation is almost the same approach as the 3-2-3 model already explained.
	Therefore we show only group $S_Z$ consisting of four nodes.
	To maximize $M_{3}M_{4}M_{7}M_{8}$ subject to
	\begin{eqnarray*}
		T=s_{3,4}+s_{3,8}+s_{4,8}+s_{3,9}+s_{4,9}+ \cdots+s_{3,10},
		\label{T3478_m224p1}
	\end{eqnarray*}
	we need to solve two equations independently
	\begin{itemize}
		\item
		Case 1: by ignoring $s_{3,4}$ (if $q_{10} \geq q_4$),
		\begin{eqnarray}
		T=2s_{3,8}+3s_{3,9}+4s_{3,10},
		\quad
		s'_{8,10}=s_{3,10},
		\quad s_{3,4}=s'_{8,10}
		\label{T3478_m224c1p1}
		\end{eqnarray}
		
		\item
		Case 2: by ignoring $s_{8,10}$ (if $q_{10} < q_4$),
		\begin{eqnarray}
		T=s_{3,4}+2s_{3,8}+3s_{3,9}+3s_{3,10},
		\quad
		s'_{8,10}=s_{3,4} 
		\label{T3478_m224c2p1}
		\end{eqnarray}
	\end{itemize}
	
	By solving Eq.~(\ref{T3478_m224c1p1}) or Eq.~(\ref{T3478_m224c2p1}), 
	a natural number solution for $S_Z$ is finally obtained.
	Based on the final natural number solution, $s_{3,4}^* + 2s_{3,8}^*$ is used
	to solve group $S_Y$.

	\subsection{Static slot allocation for 2-1-5 model}
	
	\begin{figure}
		\centering
		\includegraphics[scale=.5]{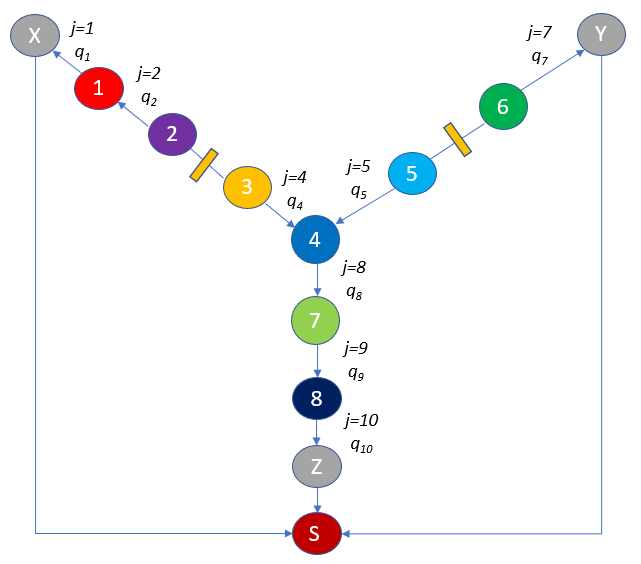}
		\caption{The 2-1-5 path model in the network topology}
		\label{jf09-m215}
	\end{figure}
	
	\begin{figure}[b]
		\centering
		\includegraphics[scale=.55]{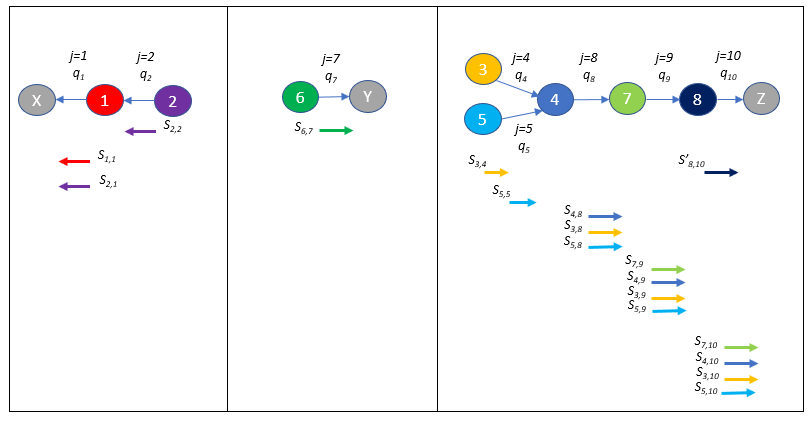}
		\caption{Transmission scheduling on 2-1-5 model (all $r_i$=1)}
		\label{jf10-m215-p}
	\end{figure}
	
	Fig.~\ref{jf09-m215} shows the 2-1-5 path model. 
	In this example, since nodes 3, 4, and 5 are assumed to be in the radio propagation distance, nodes 3, 4  and 5 cannot send at the same time (to avoid an interference). On the other hand, nodes 2 and 6 can send to its next node at the same time.
	Note that groups $S_X$ and $S_Y$ are independent and can be solved separately.
	Therefore, we need only a single pattern.

	Fig.~\ref{jf10-m215-p} shows a slot allocation on 2-1-5 model. 
	In group $S_X$,
	\begin{eqnarray*}
		M_1&=&(1-q_{1}^{s_{1,1}}),\quad
		M_2\:=\:(1-q_{1}^{s_{2,1}})(1-q_{2}^{s_{2,2}})
		\label{M_12_t251}
	\end{eqnarray*}
	In group $S_Z$,
	\begin{eqnarray*}
		M_3&=&(1-q_{4}^{s_{3,4}})(1-q_{8}^{s_{3,8}})(1-q_{9}^{s_{3,9}})(1-q_{10}^{s_{3,10}})
		\nonumber
		\\
		M_5&=&(1-q_{5}^{s_{5,5}})(1-q_{8}^{s_{5,8}})(1-q_{9}^{s_{5,9}})(1-q_{10}^{s_{5,10}})
		\nonumber
		\\		
		M_4&=&(1-q_{8}^{s_{4,8}})(1-q_{9}^{s_{4,9}})(1-q_{10}^{s_{4,10}})
		\nonumber	
		\\
		M_7&=&(1-q^{s_{7,9}}_9)(1-q^{s_{7,10}}_{10}), \quad
		M_8\:=\:(1-q^{{s_{8,10}}+{s'_{8,10}}}_{10})
		\label{M35478_m251}
	\end{eqnarray*}
	In group $S_Y$,
	\begin{eqnarray*}
		M_6\:=\:(1-q_{7}^{{s_{6,7}}})
		\label{M_6_m251}	
	\end{eqnarray*}

	The process to get a slot allocation is almost the same approach as the previous models.
	We only mention group $S_Z$.
	To maximize $M_{3}M_{5}M_{4}M_{7}M_{8}$
	subject to
	\begin{eqnarray*}
		T=s_{3,4}+s_{5,5}+s_{3,8}+s_{4,8}+s_{3,9}+ \cdots+s_{3,10},
		\label{T37456_m251p1}
	\end{eqnarray*}
	two cases should be examined and select best one.
	One case is to ignore $s_{8,10}$, i.e., node 8 does not generate own packet,
	and 
	the other case is to ignore $s_{3,4}$ and $s_{5,5}$,
	i.e., nodes 3 and 5 do not generate own packets, instead node 4 generates three packets.

	\section{Numerical results}

	On our example of the Y-shaped topology network, we show a few numerical results for three different path models to evaluate the performance of derived time-slot allocations in three different cases in terms of the setting of link loss rates $\{q_{i,j}\}$ shown in Table 1; packet generation rates are uniform ($r_i = 1$); the total number $T$ of time-slots is $T=20$ or $T=30$.
	Highly lossy links (links with high loss rates) are located near gateway $S_Z$  in case 1; near gateway $S_X$ in case 2; and near gateways $S_X$ and $S_Y$ in case 3.
	Matlab is used to get the solutions of the maximization problems for the path model in the way described in Section 4.
	As a performance metric, the Theoretical Upper-Bound (TUB) value and the Model-based Computed (COM) value are used.
	TUB is the theoretical maximum value of the objective function M(s) in the relaxed version of the maximization problem (i.e., any real number can be used).
	COM is the computed probability of delivering all packets using an optimal slot allocation according to a natural number solution of the original integer-constraint maximization problem.

	\begin{figure}[h]
		\centering
		\includegraphics[scale=.75]{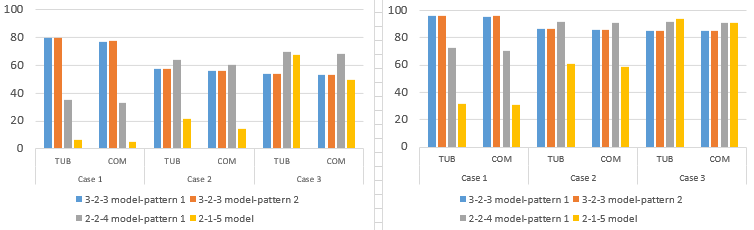}
		\caption{Probability of success delivery for all nodes with T=20 (left) and T=30 (right)}
		\label{Re-1}
	\end{figure}
	
	\begin{figure}[h]
		\centering
		\includegraphics[scale=.75]{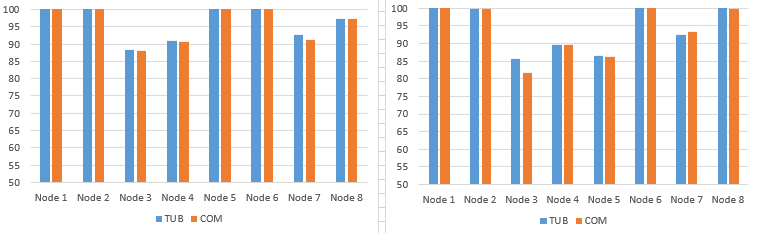}
		\caption{Probability of success delivery for each node with $T=30$, pattern 1 in the 2-2-4 model-case 1 (left), 2-1-5 model-case 1 (right) }
		\label{Re-2}
	\end{figure}
	
	In Fig.~\ref{Re-1}, the success delivery probability for all packets of TUB value performs better COM value in all cases. It is because TUB value is the theoretical maximum value of the objective function. 
	
	In all conditions, two patterns in the 3-2-3 model represent the equal TUB value regardless of whether group $S_X$ is prioritized or $S_Z$ is prioritized. It is attributed to the balanced locations of the separated links.
	
	In the case 1, the 3-2-3 mode illustrates the significantly higher performance than its counterparts due to the positive recovery after loss link around gateway Z. Noticeably, the high resilience of this model possibly stem from the relatively equal number of nodes in all groups, which relieve the effect of accumulated loss link from transmitting through many nodes. 
	When T=30, for all cases, both patterns in the 3-2-3 model show more than 80 percent of the successful delivery.

	Similarly, pattern 1 in the 2-2-4 model has demonstrated to be more effective in case that there is an impact of loss link to gateway X. It only attains the efficient probability of successful delivery of packets with no inference of loss link located at node Z. Besides, 2-2-4 model shows better performance than 2-1-5 one for both case 1 and 2. Especially, it hit the top result in case 2.
	
	The 2-1-5 model including 5 nodes in group $S_Z$ is always affected seriously by the accumulated loss link of the transmitting process towards gateway Z. In the case 3, the 2-1-5 model presents enhancement in this parameter compared with others. Our study indicates that this phenomenon only occurs in the condition that its links are not damaged by interference but links near gateway X and Y are attacked seriously. The separation link is far away from the central node leads to the arrangement of packets become more complex and vulnerable.
	
	In the simulation model with a low time-slot value (T=20), we observe the probability of successful delivery reduces significantly in all 3 cases. Herein, the effect on model 2-1-5 was the most significant because the number of time-slots assigned to the link is low. 
	
	\begin{table}[h]
		\caption{Packet loss rate on each link}
		\begin{tabular}{|c|c|c|c|c|c|c|c|c|c|c|} \hline
			Case & $q_1$ & $q_2$ & $q_3$ & $q_4$ & $q_5$ & $q_6$ & $q_7$  & $q_8$ & $q_9$& $q_{10}$ \\ \hline
			1 & 0.2 & 0.1 & 0.2 & 0.3 & 0.2 & 0.3 & 0.2 & 0.2 & 0.5 & 0.3 \\ \hline
			2 & 0.5 & 0.4 & 0.5 & 0.3 & 0.2 & 0.3 & 0.1 & 0.2 & 0.3 & 0.2 \\ \hline
			3 & 0.4 & 0.3 & 0.6 & 0.1 & 0.1 & 0.7 & 0.7 & 0.1 & 0.1 & 0.1 \\ \hline		
		\end{tabular}
		\label{tab:loss}
	\end{table}

	Fig.~\ref{Re-2} investigates the relationship between the probability of successful delivery for each node and their location in the case of 2-2-4 model in case 1 and 2-1-5 model in case 2.
	For each node, the COM value is relatively equal to the TUB value, excluding nodes under the considerable influence of the loss link. 
	In specific, the COM value of group $S_Z$ in 2-1-5 model (case 2) is higher than its TUB value at nodes 7 and 4, but this edge is reversed at nodes 3 and 8. In the same manner,  we observe a similar phenomenon in the group $S_Z$ in 2-2-4 model (case 1) but the gap between TUB and COM values is negligible.
	
	Besides, the probabilities of successful delivery for the upstream nodes are generally lower than those of the downstream nodes along a path (i.e., in a node group). The probability of successful delivery of each node is based on the number and location of nodes within a group.

	\begin{table}[h]
		\caption{Slot allocations for pattern 1 in 3-2-3 model with T=30}
		\begin{tabular}{|c|c|c|c|c|c|c|c|c|c|c|c|c|} \hline
			&$s_{1,1}$ &$s'_{1,1}$ & $s_{2,1}$& $s'_{2,1}$ & $s_{2,2}$ & $s'_{2,2}$& $s_{3,1}$ & $s_{3,2}$ & $s_{3,3}$ & $s_{4,8}$  & $s_{4,9}$ & $s_{4,10}$\\ \hline
			TUB & 5.5001 & 0 & 5.5001 & 0 & 3.9999 & 0 & 5.5001 & 3.9999 & 5.5001 & 3.4322& 6.7617 & 4.3481 \\ \hline
			COM & 5 & 0 & 5 & 0 & 4 & 0 & 6 & 4 & 6 & 3 & 7 & 4 \\ \hline
			&$s_{5,6}$ &$s_{5,7}$ & $s_{6,7}$& $s'_{6,7}$ & $s_{7,9}$ & $s'_{7,9}$& $s_{7,10}$ & $s'_{7,10}$ & $s_{8,10}$ & $s'_{8,10}$& &\\ \hline
			TUB & 11.8741 & 9.0630 & 9.0630 & 0 & 0 & 6.7617 & 3.5839 & 0.7642 & 0 & 4.3481 &  &  \\ \hline
			COM & 12 & 9 & 9 & 0 & 0 & 7 & 4 & 1 & 0 & 4&  & \\ \hline		
		\end{tabular}
		\label{P1-3-2-3-T30}
	\end{table}

	\begin{table}[h]
		\caption{Slot allocations for pattern 2 in 3-2-3 model with T=30}
		\begin{tabular}{|c|c|c|c|c|c|c|c|c|c|c|c|c|} \hline
			&$s_{1,1}$ &$s'_{1,1}$ & $s_{2,1}$& $s'_{2,1}$ & $s_{2,2}$ & $s'_{2,2}$& $s_{3,1}$ & $s_{3,2}$ & $s_{3,3}$ & $s_{4,8}$  & $s_{4,9}$ & $s_{4,10}$\\ \hline
			TUB & 5.5001 & 0 & 5.5001 & 0 & 0.5677 & 3.4322 & 5.5001 & 3.9999 & 5.5001 & 3.4322& 6.7617 & 4.3481 \\ \hline
			COM & 5 & 0 & 5 & 0 & 1 & 4 & 5 & 4 & 6 & 4 & 7 & 4 \\ \hline
			&$s_{5,6}$ &$s_{5,7}$ & $s_{6,7}$& $s'_{6,7}$ & $s_{7,9}$ & $s'_{7,9}$& $s_{7,10}$ & $s'_{7,10}$ & $s_{8,10}$ & $s'_{8,10}$& &\\ \hline
			TUB & 11.8741 & 9.0630 & 5.6307 & 3.4322 & 6.7617 & 0 & 4.3481 & 0 & 4.3481 & 0 &  &  \\ \hline
			COM & 12 & 9 & 5 & 4 & 7 & 0 & 4 & 0 & 4 & 0&  & \\ \hline		
		\end{tabular}
		\label{P2-3-2-3-T30}
	\end{table}

	Table ~\ref{P1-3-2-3-T30} and table ~\ref{P2-3-2-3-T30} show slot allocations for pattern 1 and pattern 2 in the 3-2-3 model with T=30. 
	In pattern 1, we prioritize the transmission in groups $S_X$ (node 3-2-1) and $S_Y$ (node 5-6) first, then group $S_Z$ (node 4-7-8). In pattern 2, we prioritize group $S_Z$ first, then groups $S_X$ and $S_Y$.
	
	Due to a difference of pattern, we recognized the gap in the number of time-slots distributes to each node. 
	For instance, at node 7 of pattern 2 in the 3-2-3 model, time-slots were assigned to $s_{7,9}$, whereas the counterparts in pattern 1 in the 3-2-3 model were located at $s'_{7,9}$.
	
	It is because node 4 cannot send at the same time as node 3 and node 5 in pattern 1 in the 3-2-3 model. Therefore, node 7 was prioritized to transmit simultaneously with node 3 and node 4. Otherwise, the number of time-slots allocated to $s'_{7,9}$, $s'_{7,10}$, $s'_{8,10}$ would depend on $s_{3,3}$ or $s_{5,6}$. 
	Hence, we utilized Model-based Comupted (COM) to obtained $s'_{7,9}$, $s'_{7,10}$, $s'_{8,10}$ with value at 7,1, 4, respectively, in condition that $s_{5,6}$ is equal to 12.
	
	Otherwise, from results of Table ~\ref{P1-3-2-3-T30} and table ~\ref{P2-3-2-3-T30}, the change of prioritized transmitting in each node group would facilitate to avoid interference between different nodes in the same group. However, the performance was the same in patterns 1 and 2, since the number of time-slots assigned to each node was large.

	\section{Discussion}
	We address a few issues that are not well mentioned in the main part of this paper but necessary to implement and extend our proposed scheme into practical systems.
	Firstly, our scheme is applicable to heterogeneous packet generation rates of nodes and heterogeneous data transmission rates of links, although only homogeneous cases were explained for concise formulations.
	In reality, each node may support different numbers of and/or types of sensors, and thus the number of packets necessary to convey them in one cycle period may differ.
	In addition, different types of backbone links may be mixed in the same network with the different data transmission rate to adjust some restrictions, e.g., physical distance and cost.
	In our preliminary work on the tandemly-arranged topology networks with two gateways, we have shown the formulations and results on heterogeneous packet generation rates (\cite{Yoshida20}) and data transmission rates (\cite{Kimura20}).
	Note that how the different packet generation rates ($r_i$ for node $i$) can be managed in our setting is shown in Appendix.
	Furthermore, the number of nodes can be extended, although only 8-node example was shown.
	In general, even if the number of nodes, i.e., the length of a path, is increased, the interference patterns around the central node are unchanged.
	Only the chances of a concurrent use of the same time-slots for the same direction transmissions by two distanced nodes in the path is increased.
	However, an investigation on performance impacts and implications to scheduling design remains as future work.

	Secondly, in this paper, our scheme only adopts a basic redundant transmissions in which a node just redundantly transmits each of its possessed packets in a specific times according to the given time-slot allocation. 
	However it is well-known that a packet-level coding as FEC increases the success probability of packet delivery, although it possibly increases the complexity of the system.
	Each node can combine multiple different packets its possessed by using some coding and transmits possibly different coded packets within the allocated slots; those coded packets are decoded in the final receiver, e.g., a central data collection server.
	In our preliminary work, we have shown the benefits of XOR-based simple coding with consideration on fairness among nodes (\cite{Kimura20}).
	An detailed design and performance investigation of packet-coding in Y-shaped topology setting remain as future work.

	Finally, 
	our scheme is on a centrally-managed transmission scheduling for a network of relay nodes with gateways, and implicitly assumes a central management server that compute a global time-slot allocation.
	Therefore, the next research will concentrate on the system architecture for real implementation.
	More specifically, a scheme to exchange and share the involved information is necessary (i) for a server to know or estimate a network topology and related information such as data transmission rates (bandwidths) of links, distances between nodes, packet loss rates on links, and packet generation rates at nodes; and (ii) for each node to know a derived transmission schedule.
	In particular, the information exchange of (ii) requires an opposite direction communication from gateways to each node and is needed not only at the initial phase of the system but every time when environmental conditions change or periodically with a relatively long time interval.

	\section{Concluding Remarks}

	In this paper, we focus on the linear topology with three gateways at the edges to resolve existing problems, such as interferences among relay nodes and unreliable lossy links.
	This goal would be achieved by utilizing a global static time-slot allocation to maximize the theoretical probability that all packets are successfully delivered to one of the gateways within one cycle period with redundant transmissions.
	
	Our work presented three types of the path models on Y-shaped topologies. The optimization to the packet scheduling was built on completed formulas and calculations by Matlab. All models were evaluated in different cases to select the appropriate one for each case.
	In the future, as addressed in Discussion section, we will continue to improve both the applicability and performance of the scheme, and to develop a system implement with performance evaluation based on realistic wireless communication simulations.
	
	The research results have been achieved by the ``Resilient Edge Cloud Designed Network (19304),'' NICT, and by JSPS KAKENHI JP20K11770, Japan.

	\appendix
	\section{Appendix}

	We explain how to derive Eq.(\ref{s11_t323p1}).
	For group $S_X$ in transmission pattern 1 of 3-2-3 path model,
	let $M_i$ be the theoretical probability that a single packet generated by node $i$
	is successfully delivered to gateway $X$ with the basic redundant transmission scheme.
	According to Fig.~\ref{jf04-m323-p1}, we have
	\begin{eqnarray*}
		M_1&=&(1-q_{1}^{{s_{1,1}}}),
		\:\:
		M_2\:=\:(1-q_{1}^{{s_{2,1}}})(1-q_{2}^{s_{2,2}}),
		\:\:
		M_3\:=\:(1-q_{1}^{s_{3,1}})(1-q_{2}^{s_{3,2}})(1-q_{3}^{s_{3,3}}),
	\end{eqnarray*}
	by letting $s_{i,j}$ be the number of allocated slots for one packet generated by node $i$ on link $j$.
	
	Our final goal is to find a slot allocation maximizing the theoretical probability $M$ that all packets in one cycle period are successfully delivered to gateway $X$ with the basic redundant transmission scheme.
	The exact formulation of $M$ is somewhat complicated.
	By letting $s_{i,j,k}$ be the number of allocated slots for the $k$-th packet generated by node $i$ on link $j$, this probability $M$ is
	\begin{eqnarray*}
		&&
		\prod_{j=1}^{r_1}(1-q_{1}^{{s_{1,1,j}}})
		\prod_{j=1}^{r_2}(1-q_{1}^{{s_{2,1,j}}})(1-q_{2}^{s_{2,2,j}})
		\prod_{j=1}^{r_3}(1-q_{1}^{s_{3,1,j}})(1-q_{2}^{s_{3,2,j}})(1-q_{3}^{s_{3,3,j}}).
	\end{eqnarray*}
	
	However, since we deal with a relaxation version of the maximization problem 
	to apply the Lagrangian multiplier method,
	the formulation of $M$ can be simpler in the relaxation version
	by considering $s_{i,j,k} = s_{i,j}$ for $\forall k$: 
	\begin{eqnarray}
	M=
	M(\mathbf{s})=M^{r_1}_{1}M^{r_2}_{2}M^{r_3}_{3}
	\label{app_M}
	\end{eqnarray}
	where 
	$\mathbf{s} = (s_{1,1}, s_{2,1}, s_{3,1}, s_{2,2}, s_{2,3}, s_{3,3})$ 
	and
	$s_{i,j} (> 0)$ 
	are not restricted to natural numbers.
	
	Please note that ``$s_{i,j,1} \neq s_{i,j,2}$'' may happen in the original
	maximization problem for slot allocation due to the total slot number 
	is restricted by a given $T$. 
	Therefore, as we showed in Sec.~4, we should find the exact optimal 
	natural numbers $\{s_{i,j,k} | k=1,2,\ldots,r_i\}$ after obtaining a real number solution 
	$\{s_{i,j}\}$.   
	
	The relaxation version problem can be solved as follows.
	\begin{eqnarray*}
		\max \: M
		\:\:\mbox{subject to}\:\:
		T = r_1s_{1,1} + r_2(s_{2,1} + s_{2,2}) + r_3(s_{3,1} + s_{3,2} + s_{3,3})
	\end{eqnarray*}
	where $M$ is defined in Eq.(\ref{app_M}).
	
	The corresponding Lagrangian function is:
	\begin{eqnarray*}
		L &=&
		M - \lambda(r_1s_{1,1} + r_2(s_{2,1} + s_{2,2}) + r_3(s_{3,1} + s_{3,2} + s_{3,3}) - T).
	\end{eqnarray*}
	
	First we define two notations for conciseness.
	\begin{eqnarray}
	G(q,x) \:=\:
	\frac{-q^{x} \log q}{1 - q^{x}},
	\:\:
	F(q, y) \:=\: -\frac{\log(1 - y \log q)}{\log q}
	\label{app_F}
	\end{eqnarray}
	where
	$\displaystyle
	G(q, x) = \frac{1}{y}
	\:
	\Leftrightarrow
	\:
	x = F(q, y)
	$.
	
	If $\mathbf{s}$ is a soluition within the internal region,
	$\displaystyle
	\frac{\partial L}{\partial s_{i,j}} = 0
	$ 
	should be held for every $(i ,j)$.
	Hence, by differentiating $L$ with respect to $s_{1,1}$, 
	we have
	\begin{eqnarray}
	\frac{\partial M}{\partial s_{1,1}}
	&=&
	\frac{\partial M_1}{\partial s_{1,1}}(r_1 M^{r_1-1}_1) M^{r_2}_2 M^{r_3}_3
	\nonumber
	\\
	&=&
	r_1(-q_1^{s_{1,1}} \log q_1)M^{r_1-1}_1 M^{r_2}_2 M^{r_3}_3
	=
	r_1 G(q_1, s_{1,1}) M,
	\nonumber
	\\
	\frac{\partial L}{\partial s_{1,1}}
	&=&
	r_1 G(q_1, s_{1,1}) M - r_1\lambda
	\:=\:0,
	\nonumber
	\\
	G(q_1, s_{1,1}) &=& \frac{\lambda}{M}
	\label{app_s11}
	\end{eqnarray}
	where $G$ is defined in Eq.(\ref{app_F}).
	
	In the same way, we have:
	\begin{eqnarray*}
		\frac{\partial L}{\partial s_{i,j}}
		&=&
		r_i G(q_j, s_{i,j}) M - r_i\lambda
		\:=\:0,
	\end{eqnarray*}
	and thus,
	\begin{eqnarray}
	G(q_1, s_{2,1})
	= 
	G(q_1, s_{3,1})
	=
	G(q_2, s_{2,2})
	=
	G(q_2, s_{3,2})
	=
	G(q_3, s_{3,3}) 
	\:=\: \frac{\lambda}{M}
	\label{app_sall}
	\end{eqnarray}
	
	From Eqs.(\ref{app_s11}) and (\ref{app_sall}),
	by letting 
	$\displaystyle
	\alpha = \frac{M}{\lambda}
	$
	as an adjunct variable,
	we have an explicit expression of each $s_{i,j}$ with
	an unknown positive variable $\alpha$:
	\begin{eqnarray}
	s_{1,1} = s_{2,1} = s_{3,1} = F(q_1, \alpha),
	\:\:
	s_{2,2} = s_{3,2} = F(q_2, \alpha),
	\:\:
	s_{3,3} = F(q_3, \alpha)
	\label{app_sfunc}
	\end{eqnarray}
	where $F$ is defined in Eq.(\ref{app_F}).

	\bibliographystyle{cas-model2-names}
	
	\bibliography{cas-refs}

	
	\bio{}
	Nguyen Vu Linh received the B.S. degree and M.S. degree in Electronics and Telecommunications, from University of Science, Ho Chi Minh City, Vietnam. He is currently pursuing a Ph.D. degree within the Graduate School of Computer Science and Systems Engineering, Kyushu Institute of Technology from Kyushu Institute of Technology, Japan. His research interests are network management and communication systems.
	
	\endbio
	
	\bio{}
	Nguyen Viet Ha received the B.S. degree (2009), M.S. degree (2012) in Electronics and Telecommunications, from University of Science, Ho Chi Minh City, Vietnam and Ph.D. degree (2017) in Computer Science and System Engineering from Kyushu Institute of Technology, Japan. His research interests are transport layer protocols and network coding.
	
	\endbio
	
	\bio{}
	Masahiro Shibata received B.E., M.E., and D.E. degrees in Computer Science from Osaka University, in 2012 2014, and 2017, respectively. Since 2017, he has been an Assistant Professor in Kyushu Institute of Technology. His research interests include distributed algorithms and network management. He is a member of IPSJ and IEICE.
	
	\endbio
	
	\bio{}
	Masato Tsuru received the B.E. and M.E. degrees from Kyoto University, Japan in 1983 and 1985, and then received his D.E. degree
	from Kyushu Institute of Technology, Japan in 2002. He has been a professor in the Department of Computer Science and Electronics, Kyushu Institute of Technology since 2006. His research interests include performance measurement, modeling, and management of computer communication networks. He is a member of the ACM, IEEE, IEICE, IPSJ, and JSSST.
	
	\endbio
	
\end{document}